\documentclass[12pt]{reportj}
\usepackage{deluxetablej}
\usepackage{hyperref} 
\makeatletter
\def\@to{to}
\makeatother
\usepackage{times}
\usepackage{graphicx}
\usepackage{tocloft}
\usepackage{booktabs}
\usepackage{caption}
\usepackage{xcolor}
\usepackage{amsmath,amssymb}
\usepackage{fancyheadings}
\usepackage{tablefootnote}
\usepackage{threeparttable}
\emergencystretch=\maxdimen
\usepackage{datetime}
\newdateformat{ddmonthyyyy}{\THEDAY\ \monthname[\THEMONTH]\ \THEYEAR}

\renewcommand\thesection{\arabic{section}}
\renewcommand\thesubsection{\thesection.\arabic{subsection}}

\def\ssection#1{\setcounter{subsection}{0} \refstepcounter{section} \section*{\hbox to \hsize{\large\bf \arabic{section}. #1\hfill }}\label{sec} \addcontentsline{toc}{section}{\arabic{section}. #1}}
\def\ssubsection#1{\setcounter{subsubsection}{0} \refstepcounter{subsection}\subsection*{\hbox to \hsize{\normalsize\bfseries\itshape \arabic{section}.\arabic{subsection} #1\hfill}}\label{subsec} \addcontentsline{toc}{subsection}{\arabic{section}.\arabic{subsection} #1}}
\def\ssubsubsection#1{\refstepcounter{subsubsection}\subsection*{\hbox to \hsize{\normalsize\it \arabic{section}.\arabic{subsection}.\arabic{subsubsection} #1\hfill}}\label{subsubsec} \addcontentsline{toc}{subsubsection}{\arabic{section}.\arabic{subsection}.\arabic{subsubsection} #1}}

\def\ssectionstar#1{\section*{\hbox to \hsize{\large\bf #1\hfill}} \addcontentsline{toc}{section}{#1}}
\def\ssubsectionstar#1{\subsection*{\hbox to \hsize{\normalsize\bfseries\itshape #1\hfill}} \addcontentsline{toc}{subsection}{#1}}
\def\ssubsubsectionstar#1{\subsection*{\hbox to \hsize{\normalsize\it  #1\hfill}} \addcontentsline{toc}{subsection}{#1}}

\renewcommand{\cftaftertoctitle}{%
\mbox{}\hfill{\normalfont Page}}
\begin{document}

~\\

\vspace{-2.4cm}
\noindent\includegraphics*[width=0.295\linewidth]{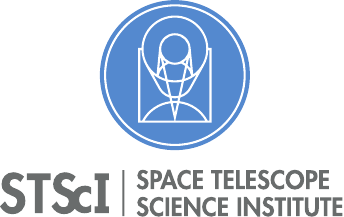}

\vspace{-0.4cm}

\begin{flushright}
    {\bf Instrument Science Report COS 2024-01(v2)}
    
    \vspace{1.1cm}
    
    {\bf\Huge The Hubble Advanced Spectral Product (HASP) Program\\}
  
    \rule{0.25\linewidth}{0.5pt}
    
    \vspace{0.5cm}
    
    John Debes$^{1,2}$, Ravi Sankrit$^1$, Travis Fischer$^{1,2}$, Elaine Frazer$^1$, Alec Hirschauer$^1$, Kate Rowlands$^{1,2}$, Matthew Burger$^1$, Robert Swaters$^1$, Robert Jedrzejewski$^1$, Sierra Gomez$^1$, Leonardo Dos Santos$^1$, Svea Hernandez$^{1,2}$, Lauren Miller$^1$, Anna Payne$^1$, Marc Rafelski$^1$, Thomas Wevers$^1$, Sara Anderson$^1$, Tom Bair$^1$, Kathryn Bello$^1$, Joleen Carlberg$^1$, Brian Charlow$^1$, Andrew Cortese$^1$, Nadia Dencheva$^1$,  Tracy Ellis$^1$, Ben Falk$^1$, Scott Fleming$^1$, Peter Forshay$^1$, Syed Gilani$^1$, Patty Hall$^1$, Tim Kimball$^1$, Talya Kelley$^1$, Richard Kidwell$^1$, Jenn Kotler$^1$, Aiden Kovacs$^1$, Bethan James$^{1,2}$, Christopher Rahmani$^1$, David Rodriguez$^1$, Julia Roman-Duval$^1$, David Soderblom$^1$, Lisa Sherbert$^1$, Dan Welty$^1$, David Wolfe$^1$
    \linebreak
    \newline
    \footnotesize{$^1$ Space Telescope Science Institute, Baltimore, MD\\}
    \footnotesize{$^2$ AURA for the European Space Agency\\
                     	}
    
    \vspace{0.4cm}
    
     16 January 2024
\end{flushright}

\noindent\rule{\linewidth}{1.0pt}
\noindent{\bf A{\footnotesize BSTRACT}}

{\it \noindent The Hubble Advanced Spectral Products (HASP) program is designed to robustly coadd Cosmic Origins Spectrograph (COS) and Space Telescope Imaging Spectrograph (STIS) spectra within the Mikulski Archive for Space Telescopes (MAST) in an automated fashion such that coadds are available for new data or archival data with updated calibrations. For each target within a visit or program, HASP employs a meticulous multi-stage filtering process to ensure data quality and creates coadded products for all central wavelengths (CENWAVEs) within specific gratings, as well as combined products using different gratings and instruments. The project also emphasizes making the code accessible to the user community for custom coaddition. As calibrations improve and new data are added to the archive, HASP products are re-created automatically so that they represent the best reduction of a given visit or program. Automated coadditions like those achieved by HASP can significantly enhance the combination of different CENWAVES, increase signal-to-noise ratios, and increase wavelength coverage. These properties make HASP a vital resource for astronomers using archival spectroscopic data from HST.}
\vspace{-0.1cm}
\noindent\rule{\linewidth}{1.0pt}

\renewcommand{\cftaftertoctitle}{\thispagestyle{fancy}}
\tableofcontents


\vspace{-0.7cm}
\ssection{Introduction}\label{sec:Introduction}
The ultimate goal of the Hubble Advanced Spectral Products, or HASP, program is to ease access to HST spectroscopic data by providing high quality one-dimensional (1-D) spectra that are robustly and flexibly combined. With data products that are automatically updated as calibrations change, a user can immediately access spectroscopic data in a meaningful way without having to resort to re-inventing spectroscopic coaddition.


\lhead{}
\rhead{}
\cfoot{\rm {\hspace{-1.9cm} Instrument Science Report COS 2024-01(v2) Page \thepage}}

The default pipeline products of COS and STIS typically do not include combinations of different gratings, central wavelengths (CENWAVEs), or apertures for individual programs. This has historically meant that users were responsible for coadding different spectra for final scientific analyses. The construction of the Hubble Spectroscopic Legacy Archive (Peeples et al., 2017; HSLA) involved coadding multiple COS modes into single 1-D spectra. This algorithm was updated and used for the UV Legacy Library of Young Stars as Essential Standards (ULLYSES) program in order to create high level science products of spectra from a variety of modes and spectrographic instruments (Roman-Duval, et al. 2020).

The HASP coadding algorithm is designed to extend beyond both previous efforts so that it can be applied in an automated fashion to nearly every COS and STIS spectrum in the archive, creating robust coadded and combined products for programs and their constituent visits. Additionally, the HASP project also enables custom coaddition to the community through a publicly available Python script and accompanying Jupyter notebooks. HASP provides a foundation for automated multi-program coaddition in order to create an updated HSLA in the future. 

The HASP products are designed to work best for astrophysical sources that have compact angular sizes and are not variable, which ensures flux continuity across different gratings and apertures. In those cases, the coadditions should match each instrument's requirements for flux and wavelength repeatability and accuracy. 

However, HASP does not fit all science cases or requirements, so an integral part of the project is to provide user-friendly access to custom coaddition for situations when more individualized approaches might need to be taken. HASP custom coaddition opens the possibility that future archival analysis projects might build on the HASP legacy in order to create high level science products that are consistent with both HASP and ULLYSES.

This ISR describes the details of the process by which default 1-D spectra in the archive are combined into final HASP coadds. Section 2 describes the scientific requirements for HASP and the process of dataset filtering and coaddition methodology for visit- and program-level coadds. Section 3 gives an overview of the data product structure, and Section 4 describes the testing and scientific validation that was performed on the HASP coadds. In Section 5 we briefly describe how users can access the Jupyter notebooks that provide examples of how to create custom coadds. Section 6 includes a set of caveats with respect to the default HASP coadds and a list of the types of datasets that will not generate default products.  In the concluding Section 7, we briefly summarize the HASP project achievements and outline prospects for the future. Appendix A provides further details of how spectra from different gratings or instruments are joined together, or abutted, Appendix B describes HASP data product BINTABLE extensions, and finally Appendix C provides an example of a default HASP output log that is delivered with HASP coadds.

\ssection{Visit and Program Coaddition}\label{sec:coadd}

Default HASP coadds are created through a three-step process in the MAST archive--automated collection of datasets, dataset filtering and validation, and coaddition. At the end of coaddition, each HASP product must adhere to the overall scientific requirements laid out by the project to ensure that the coadds are viable for scientific analysis.

\subsection{HASP Scientific Requirements}\label{sec:req}

The HASP scientific requirements are that $>75$\% of coadds should have flux and wavelength accuracy within 5\% of the input 1-D spectra. Additionally, the HASP coadded products of known flux standard stars were required to meet the formal flux and wavelength accuracies of their corresponding modes (see Section~\ref{sec:test} for more details). The requirements are for non-variable isolated and compact sources ($r_\mathrm{source} \lesssim 0.3$'') with successful 1-D extractions (i.e., x1d or sx1 products). HASP pre-filters failed or unsupported observations to meet the above requirements (see Section~\ref{filtering} for more details). Extended and variable sources are still coadded, but are considered available and unsupported, so may not meet instrument requirements for a given mode or across different gratings or apertures.

\subsection{Collation of Input Spectra}

Initial queries of the MAST databases are executed by an automated workflow that is responsible for collecting relevant datasets for processing and defines which input spectra are selected for coaddition. This process works on both public and proprietary data, such that PIs can access HASP products as well as users who wish to access public HASP products.

The workflow checks for datasets that have updated default 1-D extracted products (i.e., x1d or sx1 FITS files). These updates could be driven by recently executed observations, or by reprocessing of existing observations due to new instrument calibrations. There is logic in the workflow to check that, in the case of new observations, the entire new visit has completed processing and that there are no new visits in the near future. For reprocessed data, the workflow checks that there are no more datasets from the program in the reprocessing queue before the HASP coadditon begins. This ensures that the program is not processed repeatedly. Checks are done to make sure that proprietary data is not mixed with public data.

The HASP workflow copies all input spectra into a local processing directory, excluding spectra with archive data quality severity codes that indicate a possible failed observation, and calls the HASP wrapper. The wrapper is responsible for calling the \texttt{coadd} code, conducting supplementary quality checks (See Section \ref{filtering}), verifying flux values (See Section \ref{fluxchecking}), and generating the final data products. The HASP code determines how the files should be associated in a visit and program using the target name keyword in the FITS header.

\subsection{Filtering and Validation of Input Spectra}
Before the data can be coadded, a crucial step ensures that all the input data are of sufficient quality and reliability. One of the main risks to meeting the requirements lies in the presence of observatory-level target acquisition failures, which can result in data of varying quality levels. Early assessments revealed that while such failures historically account for only 10\% of observations, nearly 30\% of programs are impacted by their occurrence. Given that a significant majority of these programs undergo Hubble Observation Problem Reports (HOPRs) and involve repeat observations, it becomes essential for the automated coaddition process to identify and filter out any potentially spurious data. A second check of flux levels across input spectra allows any other failures or variability (mis-labeling of guide star failures, PI error, or incorrect spectral extraction) to be removed.

\subsubsection{Filtering Input Spectra}
\label{filtering}
To address this challenge, the HASP coaddition procedure employs a multi-stage filtering process. First, it compiles a list of all exposures associated with AlertObs reports and excludes those exposures. COS and STIS special calibration programs containing engineering commands, for example focus changes, or non-standard detector positions or modes are removed, as are datasets which are statically archived and do not undergo routine re-processing. It then removes any input spectra with header keywords that indicate data in a given program may have observing quality issues (often as a result of a failed or delayed guide star acquisition):

\begin{enumerate}
\item[] EXPFLAG: The EXPFLAG keyword is in the science extension(s) of all files. If the value is not ‘NORMAL’ due to a guide star failure or loss of target lock, or if the keyword is absent, the file is removed from the list.

\item[] EXPTIME: The EXPTIME keyword is in the science extension(s) of all files. If the value is not $>$ 0.0, this indicates that the shutter did not open, and the file is removed from the list.

\item[] PLANTIME: The PLANTIME keyword is present in the science extension(s) of COS files. If the value of the EXPTIME keyword is $<$ 80\% of the value of the PLANTIME keyword, the file is removed from the list, since this indicates guiding issues and the source may have drifted significantly in the aperture. If the PLANTIME keyword is absent, the file is processed normally (as is the case for STIS exposures). 

\item[] FGSLOCK: The FGSLOCK keyword is present in the science extension(s) of all files. If the value of the FGSLOCK keyword is not ‘FINE’, this indicates guiding issues and the file is removed from the list. 

\end{enumerate}

Observing parameters are checked and input spectra are filtered for certain cases that make accurate coadds possible:
\begin{enumerate}
\item[] POSTARG1: The POSTARG1 keyword is present in the primary header of all files. If the value of the POSTARG1 keyword is not 0.0, it is likely that observations were taken in a non-standard fashion or on a secondary target. In most cases this might mean the flux or wavelength scales may not meet requirements and the file is removed from the list. 

\item[] POSTARG2: The POSTARG2 keyword is present in the primary header of all files. If the value of the POSTARG2 keyword is not 0.0 and the pattern P1\_PURPS is not 'DITHER', the file is removed from the list under the assumption that non-standard observations were taken that might compromise flux or wavelength accuracy for the primary target. 

\item[] PATTERN1: This is to exclude STIS exposures that are part of a pattern to dither perpendicular to the slit.  Unlike dithering parallel to the slit, the pipeline cannot re-center the spectrum extraction box, so the pointings will necessarily be different.  This situation is covered by removing from the list any exposures that have the keyword PATTERN1 from the primary header equal to ‘STIS-PERP-TO-SLIT’ and the keyword P1\_FRAME from the primary header equal to ‘POS-TARG’. Files lacking these keywords are unaffected. 

\item[] P1\_PURPS: This keyword is in the primary header of STIS files.  If the value of this keyword is ‘MOSAIC’, the file is removed from the list. Files lacking this keyword are unaffected. 

\item[] OPT\_ELEM='PRISM': The STIS PRISM is an unsupported mode; PRISM data is removed from any input list.

\item[] APERTURE='BOA': The COS BOA is an unsupported mode; BOA data is removed from any input list.

\item[] COS NUV G230L Stripe C observations are not included in the coadd products as default because most of the Stripe C flux is second-order light from the Stripe A wavelength range.

\end{enumerate}

Only visit-level products are produced for moving targets (i.e., Solar system objects). Targets with the MTFLAG set to true do not have coadded products generated at the program level due to their expected variability from visit to visit. If users wish to include any of the above types of data they may use the publicly available HASP \texttt{coadd} code with keyword-based filtering disabled, except for STIS PRISM data, which is always filtered.

\subsubsection{Checking the Flux of Input Spectra}
\label{fluxchecking}
Despite several different checks for input spectra that may have been part of a failed observation, more subtle issues with target acquisition, spectroscopic extraction, or jitter may produce a default archival product that passes the HASP pre-filtering steps, but may still be inappropriate for coaddition.
To handle such cases, the HASP wrapper code implements a flux checking step that compares input spectra from a single mode against that mode's final coadd.
If the flux levels for an input spectrum are found to be lower than that of the coadd, the wrapper code will reject it, and an iterative process commences whereby a new coadd is constructed from those spectra which were retained.
The flux levels of the remaining input spectra are compared to this new coadd, and the process repeats until no additional input spectra are rejected from the next iteration.

In order to accomplish this, the flux checking algorithm determines the median deviation from the coadd ($F_{\mathrm{x1d}}(i)-F_{\mathrm{coadd}}(i)$), weighted by the uncertainty per wavelength bin of the input spectrum ($\sigma_{\mathrm {x1d}}$), and rejects an input spectrum if the following criterion is met:
\vspace{-0.3cm}
\begin{equation}
\left<\frac{F_{\mathrm{x1d}}(i)-F_{\mathrm{coadd}}(i)}{\sigma_{\mathrm{ x1d}}(i)}\right> < \frac{C_{\mathrm{thresh}}}{\sqrt{N_{\mathrm{pix}}}}
\end{equation}
where $N_{\mathrm pix}$ refers to the number of wavelength bins for a given mode.
For STIS and COS NUV this is roughly 1024 (relating to the number of pixels on the detctor), while for COS FUV this is closer to 14,000. The value for $\sigma_{\mathrm x1d}$ is defined to be the input spectrum flux divided by the signal-to-noise ratio (SNR) per wavelength bin. For high SNR datasets, the uncertainty is capped at 5\% of the flux (i.e. SNR=20/wavelength bin) to ensure that flux checking is not influenced by any systematic uncertainties in a given mode that may exceed the estimated photon-limited SNR. It is important to note here that the wavelength bin sizes may be smaller than a resolution element for a given mode. Users are encouraged to check the COS and STIS instrument handbooks to compare HASP coadd wavelength bins against the reported resolution element sizes for particular gratings.

For default products delivered by the archive, the threshold (C$_{\mathrm thresh}$) is set to -50, which has been determined to effectively reject most low-flux spectra, while retaining spectra of sufficient flux, for the majority of datasets in our test sample.
In the publicly available HASP \texttt{coadd} package, however, this threshold can be adjusted by the user to suit the needs of their science program goals. More details are available in Section \ref{sec:custom}.

\vspace{-0.4cm}
\subsection{HASP Coaddition}
After filtering, the HASP wrapper code organizes COS and STIS spectroscopic data in order to be processed by the \texttt{coadd} code. It works on all unfiltered x1d and sx1 files in the selected directory within the HASP workflow. Both COS and STIS have x1d files for UV observations, while STIS has sx1 files, derived from CCD observations subjected to CR-SPLITs. The wrapper code works on all spectroscopic files for a single program. 

The files are then sorted into groups that have the same target name (keyword TARGNAME),
grating (keyword OPT\_ELEM) and instrument (keyword\ INSTRUME).  These groups are further
split into visit-level groups (all from the same visit) and program-level groups (all
from the same program), across both gratings and instruments.  The data from each of these groups is combined to create
visit-level products and program-level products.

\subsubsection{Coaddition Methodology}

The basic coaddition procedure for a single grating descends from the ULLYSES \texttt{coadd}\footnote{\href{https://pypi.org/project/ullyses/}{https://pypi.org/project/ullyses/}} code and the coaddition methodology detailed in Peeples et al., (2017). The HASP wrapper code implements some differences in methodology with ULLYSES while maintaining backwards capability with the core \texttt{coadd} code.
 
First, maximum and minimum values from the wavelength arrays of each input exposure are
calculated, and these are rounded down (minimum wavelength) and up (maximum wavelength) to the nearest integer Angstrom.  The largest wavelength bin-size between neighboring input wavelength arrays is used as the wavelength spacing between samples of the product spectrum. 

In the \texttt{coadd} procedure, each input pixel's measurement is treated as an estimate of the monochromatic flux at its assigned wavelength. The output flux is obtained by calculating a weighted average of all the flux measurements falling within the output wavelength bin's bounds.  Throughput, which is the net count rate divided by flux and multiplied by the exposure duration, serves as the weighting factor for each input pixel.  This ensures that measurements with more counts have higher weights. Only input pixels with Data Quality (DQ) flags for COS\footnote{\href{https://hst-docs.stsci.edu/cosdhb/chapter-2-cos-data-files/2-7-error-and-data-quality-arrays\#id-2.7ErrorandDataQualityArrays-2.7.2DataQualityFlags}{COS Data Handbook}} and STIS\footnote{\href{https://hst-docs.stsci.edu/stisdhb/chapter-2-stis-data-structure/2-5-error-and-data-quality-array\#id-2.5ErrorandDataQualityArray-2.5.2DataQualityFlagging}{STIS Data Handbook}} considered non-serious contribute to the output flux.  One exception is the DQ flag of 16 (which relates to flagging of pixels with dark rates $>$5$\sigma$ times the median dark level) in STIS data products, which was found to be significantly over-flagged in current STIS data products. Errors are calculated from the square root of the total detected counts in each bin for data from
photon counting modes (COS, STIS MAMA), and from the weighted sum of the squared errors for STIS CCD data.  The conversion from error counts to error flux in the former case uses the conversion
calculated for the flux measurement at each wavelength. This procedure ignores any systematic uncertainties (such as those due to CCD fringing beyond 7000~\AA\ and pixel-to-pixel variations in COS FUV spectra).

In cases where net counts and flux are zero, the conversion ratio is interpolated using neighboring values.  The SNR is calculated for each wavelength bin as the ratio of the flux to the error. This method of combination prevents correlated errors in neighboring pixels, albeit with a minor loss in spectral sampling.

The total exposure time of input samples that contribute to a product sample is recorded as the effective exposure time. Some bins in the product spectrum do not receive any flux from the input spectra (either because some input spectra do not reach those wavelengths or the input data have DQ flags that exclude them). These bins will have an effective exposure time of 0 (and a flux of 0).

\begin{figure*}[!htbp]
\centering
\includegraphics[width=0.98\textwidth]{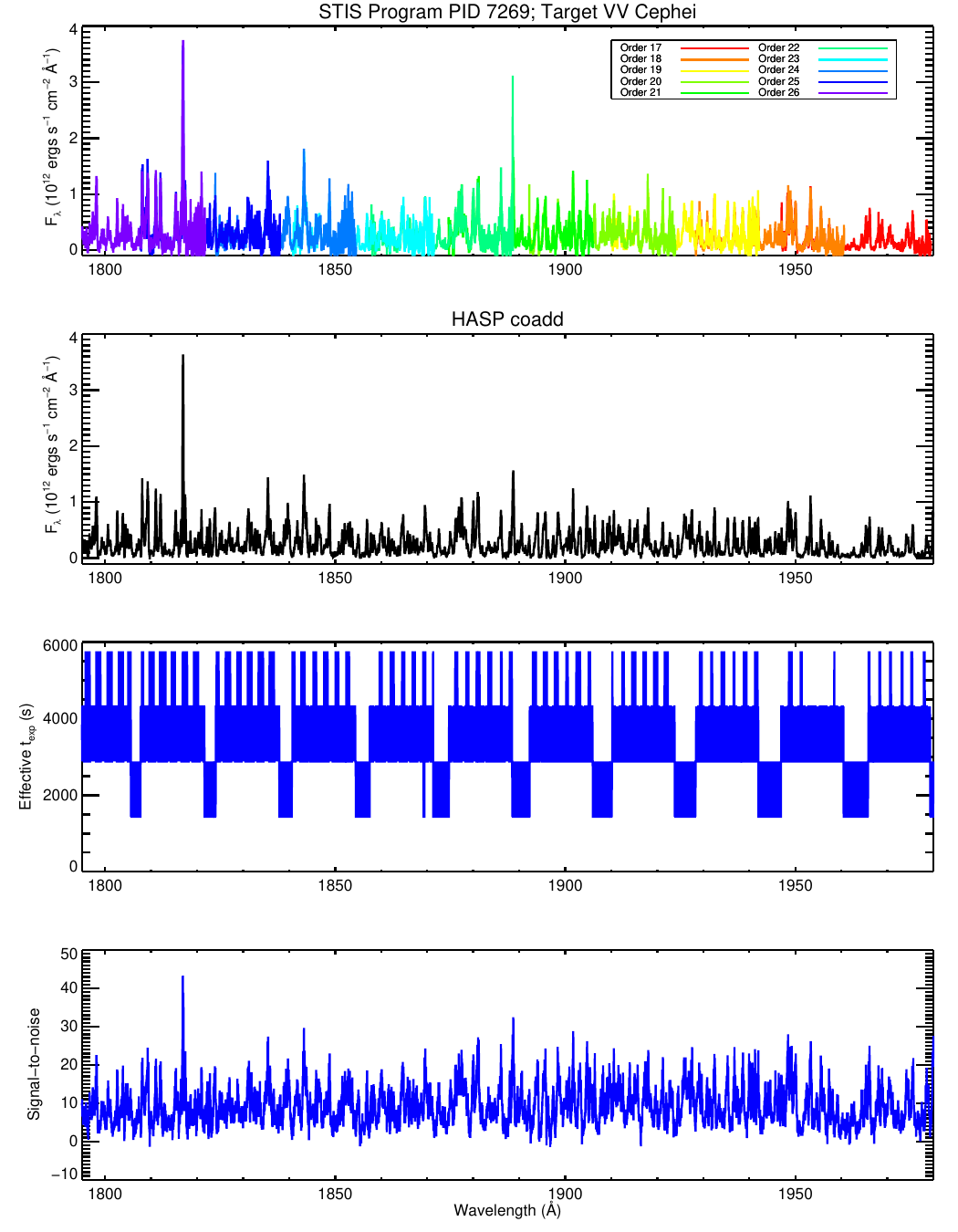}
\vspace{-0.3cm}
\caption{Example of a STIS coaddition of multiple orders from a STIS/E230M observation of VV Cephei in total eclipse, where a hot stellar companion is illuminating the wind of an M supergiant (Bauer, Bennet \& Brown, 2007). The top panel shows a subset of orders in the STIS/E230M observation. The second panel shows the resulting HASP coadded spectrum, while the third and fourth panels show the effective exposure times and SNRs, respectively, at each wavelength bin of the coadded spectrum. The spectra are shown as full resolution.}
\label{fig:coaddechelle}
\end{figure*} 

\subsubsection{Coadding Spectra from the Same Grating}

The main scientific products for each target in a program are the visit level and program level coadds, which consist of a weighted sum of spectra obtained using a common grating.
When creating coadds, two scenarios are handled that are separate from archival products: (i) combining adjacent spectral orders within a single echelle exposure and (ii) different exposures obtained with a common grating but with the same or different central wavelength settings.  

Combining echelle orders in STIS data is a particular area where HASP coadds greatly lower the barrier to accessing the data. Since each order has a wavelength overlap with another order, combining orders in a way that accurately handles the variation in sensitivity across two orders can be difficult. Now, with HASP coadds, this is handled automatically, creating seamless 1-D spectra out of a single echelle observation. An example coadd of echelle data where multiple orders are combined is shown in Figure \ref{fig:coaddechelle}.

The figure shows the input orders in the top panel and the HASP coadd in the second panel. The third panel shows the effective exposure time within each wavelength bin. In Figure \ref{fig:coaddechelle} this results in a sawtooth pattern as some wavelength bins only have flux from a single order, others have from two orders, and a subset of wavelength bins have contributions from two orders and a neighboring flux bin due to aliasing between the wavelength bin size and the input wavelength binsize.

\begin{figure*}[!htb]
\centering
\includegraphics[width=\textwidth]{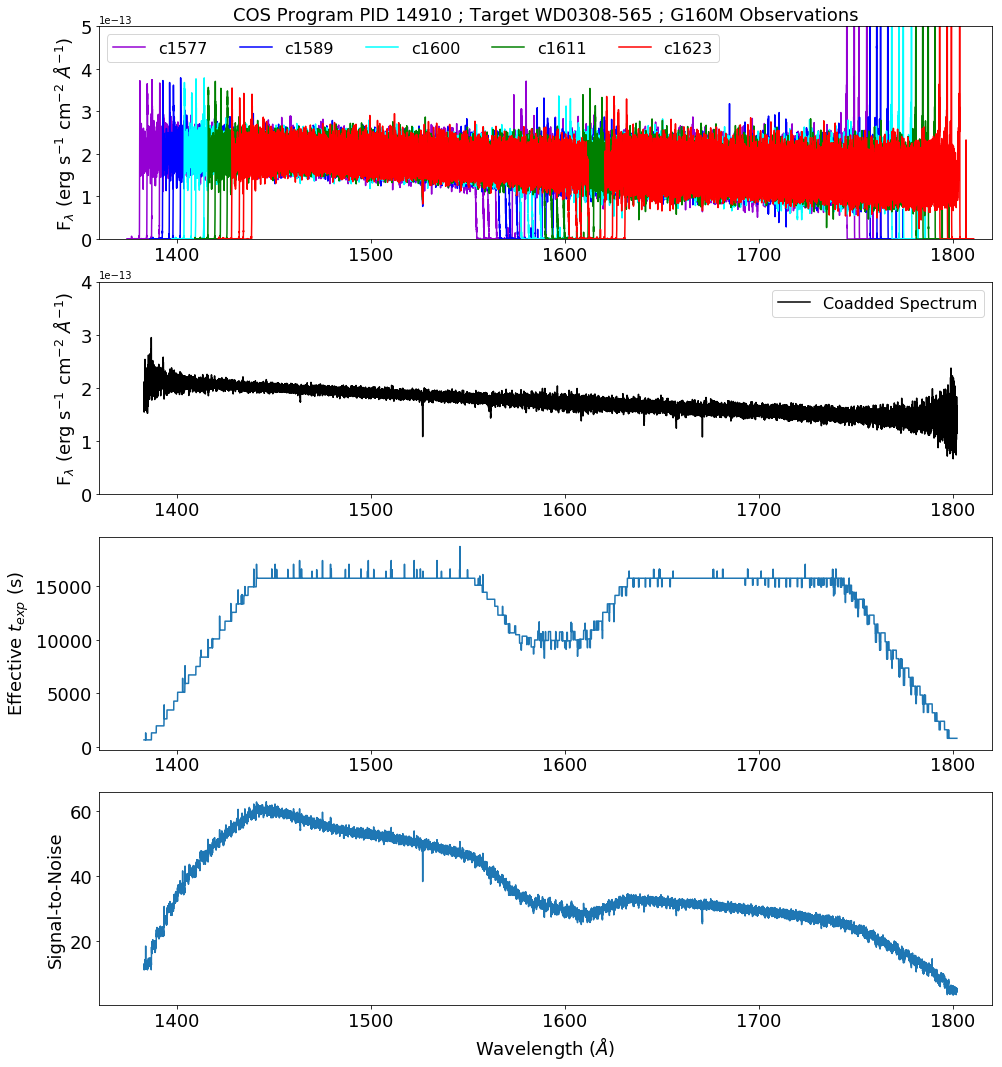}
\vspace{-0.3cm}
\caption{Example of a COS HASP coaddition of all cenwaves from the G160M grating used to observe the standard white-dwarf star WD~0308-565 as part of the LP4 flux calibration program (PID:14910, PI: Rafelski). The top panel includes 20 observations using each of five cenwaves (1577, 1589, 1600, 1611 and 1623) at all four FP-POS.  The second panel is the coadded spectrum, and the third and fourth panels show the effective exposure times and SNRs, respectively, at each wavelength bin of the coadded spectrum. The large number of CENWAVES ensure complete wavelength coverage between 1400-1800~\AA, and a significantly higher SNR. All spectra are shown at full resolution.}
\label{fig:coaddcenwave}
\end{figure*}

Many programs will obtain observations across different CENWAVES within a grating in order to increase wavelength coverage, or in the case of the COS FUV, to help minimize the degradation of the FUV detector via the rules designed to extend its lifetime (Oliveira et al., 2018). For example, observations combining G130M/1291 and G130M/1222 are recommended for full wavelength coverage in this grating. However, the default spectroscopic products do not combine CENWAVES within a grating. HASP now provides automatic CENWAVE combination, 
and an example where several cenwaves from the same COS grating are combined is shown in Figure~\ref{fig:coaddcenwave}.

\subsubsection{Joining Spectra from Different Gratings}
\label{abutment}
 A joined spectrum from all instruments and modes used in a visit or program is optimal for a first look at an object's spectral energy distribution. However, many modes have significantly different spectral resolution, making it hard to execute coaddition in the overlap regions between gratings. 

To create a spectrum using all of the coadded products from the different gratings, the spectra are abutted - i.e., joined at a single wavelength so that all the data from wavelengths below this come from one grating, and all data from wavelengths above from a different grating.  Since many of these gratings overlap, a priority list was created that governed which grating data were selected at each wavelength. 

\clearpage

For spectra with different gratings and instruments, a splicing approach is employed.
Non-overlapping spectra are abutted by aligning the arrays, resulting in a discontinuous spectral sampling at the transition.  On the other hand, for overlapping spectra, a transition wavelength is selected, and the input spectra are truncated and abutted at that wavelength, again resulting in discontinuous spectral sampling at the transition.  This approach ensures seamless combination while accommodating diverse spectral configurations.

Table~\ref{tab:abutment_priority} contains the prioritized list of gratings in order from highest priority to lowest, and for each grating the wavelength range and the number of distinct central wavelength settings available.  The priority list was designed to place a preference on spectral resolution. Exceptions to this are STIS/E140H and STIS/E230H, because these modes typically make use of slits that have a higher chance of flux variations due to the orbital focus evolution of HST.   In practice, the data from a given grating may not fill the wavelength range specified in the grating priority table. In such cases, the portion of the spectrum within this range is utilized, and the transition wavelengths are adjusted accordingly. If two gratings completely overlap a wavelength range (i.e. E140H and E140M), only the higher priority grating will be included in a coadd.

\begin{deluxetable}{cccc}
    \tabletypesize{\small}
    \tablewidth{0pt}
    \tablecaption{
    Grating Abutment Order Priorities (Highest to Lowest Priority)
    \label{tab:abutment_priority}} 

\tablehead{
    \colhead{Instrument}&\colhead{Grating}&\colhead{Wavelength Range (\AA)}&\colhead{\#Cenwaves/Tilts}}
\startdata
STIS Echelle & E140M & 1144 — 1710 & 1 \\
COS FUV & G130M & 900 — 1470 & 8 \\
COS FUV & G160M & 1342 — 1800 & 6 \\
STIS Echelle & E140H & 1140 — 1700 & 3 \\
STIS MAMA & G140M & 1140 —1740 & 12 \\
STIS Echelle & E230M & 1605 — 3110 & 2 \\
STIS Echelle & E230H & 1620 — 3150 & 6 \\
STIS MAMA & G230M & 1640 — 3100 & 18 \\
COS FUV & G140L & 900 — 2150 & 3 \\
STIS CCD & G230MB & 1640 — 3190 & 11 \\
COS NUV & G185M & 1664 — 2134 & 15 \\
COS NUV & G225M & 2069 — 2526 & 13 \\
COS NUV & G285M & 2474 — 3221 & 17 \\
STIS MAMA & G140L & 1150 — 1730 & 1 \\
STIS CCD & G430M & 3020 — 5610 & 10 \\
STIS MAMA & G230L & 1570 — 3180 & 1 \\
STIS CCD & G230LB & 1680 — 3060 & 1 \\
COS NUV & G230L & 1650 — 3200 & 4 \\
STIS CCD & G750M & 5450 — 10140 & 9 \\
STIS CCD & G430L & 2900 — 5700 & 1 \\
STIS CCD & G750L & 5240 — 10270 & 1 \\
\enddata
\end{deluxetable}

\begin{figure*}[!htbp]
\centering
\includegraphics[width=0.98\textwidth]{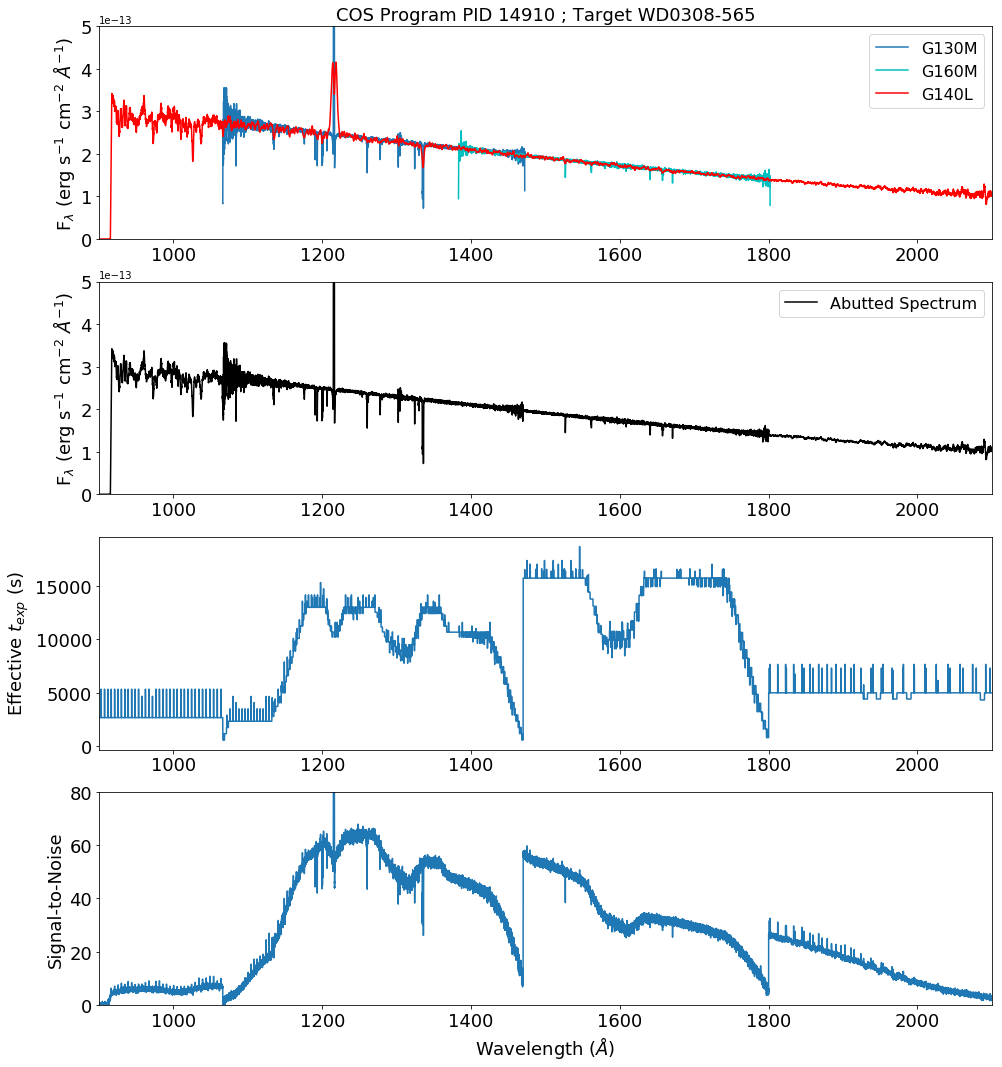}
\vspace{-0.3cm}
\caption{First panel: HASP coadded spectra of WD~0308-565 obtained as part of the LP4 Flux Calibration program. The data have been smoothed by 21 pixels for display purposes. Second panel: HASP abutted spectrum created from the coadded spectra shown in the first panel.  Third panel: the effective exposure time plotted against wavelength for the abutted spectrum.  Fourth panel: the SNR achieved in the abutted spectrum. The large drop around 1450~\AA\ and 1800~\AA\ represent differences in total exposure time at the abutment transitions between gratings.}
\label{fig:abutment_14910}
\end{figure*}

Figure~\ref{fig:abutment_14910} shows an example of how abutment works.  The data are from the COS flux calibration program (PID: 14910, PI: Rafelski) at lifetime position 4 (LP4).  In that program, the standard white dwarf star WD~0308-565 was observed through all COS modes available at LP4.  The top panel shows the coadds for each of the three gratings, the second panel shows the abutted spectrum, and the third and fourth panels show the effective exposure time and SNR in the abutted spectrum.  The spectra have been smoothed by 21 pixels for display purposes.  Note that in the abutted spectrum, G130M is prioritized over G160M, and both are prioritized over G140L.  The difference in effective exposure times is due to the exposure time per individual cenwave as well as the number of cenwaves per grating used for the observations. 

The abutted spectra are produced primarily as quick-look products, and let the user determine at a glance the overall wavelength coverage of the observation.  Since the prioritization rules are based on grating, the abutted spectra do not necessarily include the highest SNR or highest spectral resolution data in any given wavelength range.  HASP produces only visit- and program-level coadds.  Although the HASP code can handle abutment of spectra from both COS and STIS, this happens rarely at the visit-level since single visits using both COS and STIS are extremely uncommon. The use of both instruments for the same target is relatively more common within a program.

\vspace{-0.4cm}
\ssection{HASP Data Product Structure and Examples}\label{sec:dp}

The \texttt{coadd} code implemented in the HASP project generates coadd and abutment data products at both the visit and program levels. Coadded products are the result of combining spectra from a common grating, while abutments involve the combination of spectra from different gratings and/or instruments, as detailed in Section \ref{abutment}.

The naming convention for data products is similar at both the visit-level and program-level, roughly following the naming convention of the ULLYSES HLSPs\footnote{\href{https://ullyses.stsci.edu/ullyses-data-description.html}{https://ullyses.stsci.edu/ullyses-data-description.html}}. \\ 

\noindent Visit-level: \\{\small{\texttt{hst$\_$<PID>$\_$<instrument>$\_$<target>$\_$<opt$\_$elem>$\_$<ipppss>$\_$cspec.fits}}} \\

\noindent Program-level: \\
{\small{\texttt{ hst$\_$<PID>$\_$<instrument>$\_$<target>$\_$<opt$\_$elem>$\_$<ippp>$\_$cspec.fits}}} \\

In the above naming scheme, \texttt{PID} is the program ID, \texttt{instrument} is either COS or STIS or both, \texttt{target} refers to the Phase II target name, and \texttt{opt$\_$elem} is the grating name for the relevent mode(s). If a grating name is shared by COS and STIS a ``c'' or ``s'' will denote which instrument it's associated with. Finally, \texttt{ipppss} refers to the instrument, program, and dataset identifier for visit-level coadds, while \texttt{ippp} refers to the instrument and program identifier for program-level coadds. These identifiers correspond to the MAST archive dataset naming conventions. As an example, the HASP coadd shown in Figure \ref{fig:coaddechelle} is \texttt{hst\_7269\_stis\_hd208816\_e230m\_o4g201\_cspec.fits}, representing that this coadd is a visit-level coadd of just the E230M grating from Program 7269.

The HASP FITS structure for these coadded products consists of two BINTABLE extensions: A science extension containing the coadded spectra and accompanying noise and SNR estimates and a metadata extension recording attributes of each spectrum contributing to the combination, or so-called provenance table (See Tables \ref{tab:datacolumns} and \ref{tab:provenance} in Appendix B). The provenance table includes information such as the filename, telescope, instrument, detector, observation start and stop time in modified julian dates, exposure time, aperture used, and the wavelength coverage of the observation.

\begin{figure*}[!t]
\centering
\includegraphics[width=0.9\textwidth]{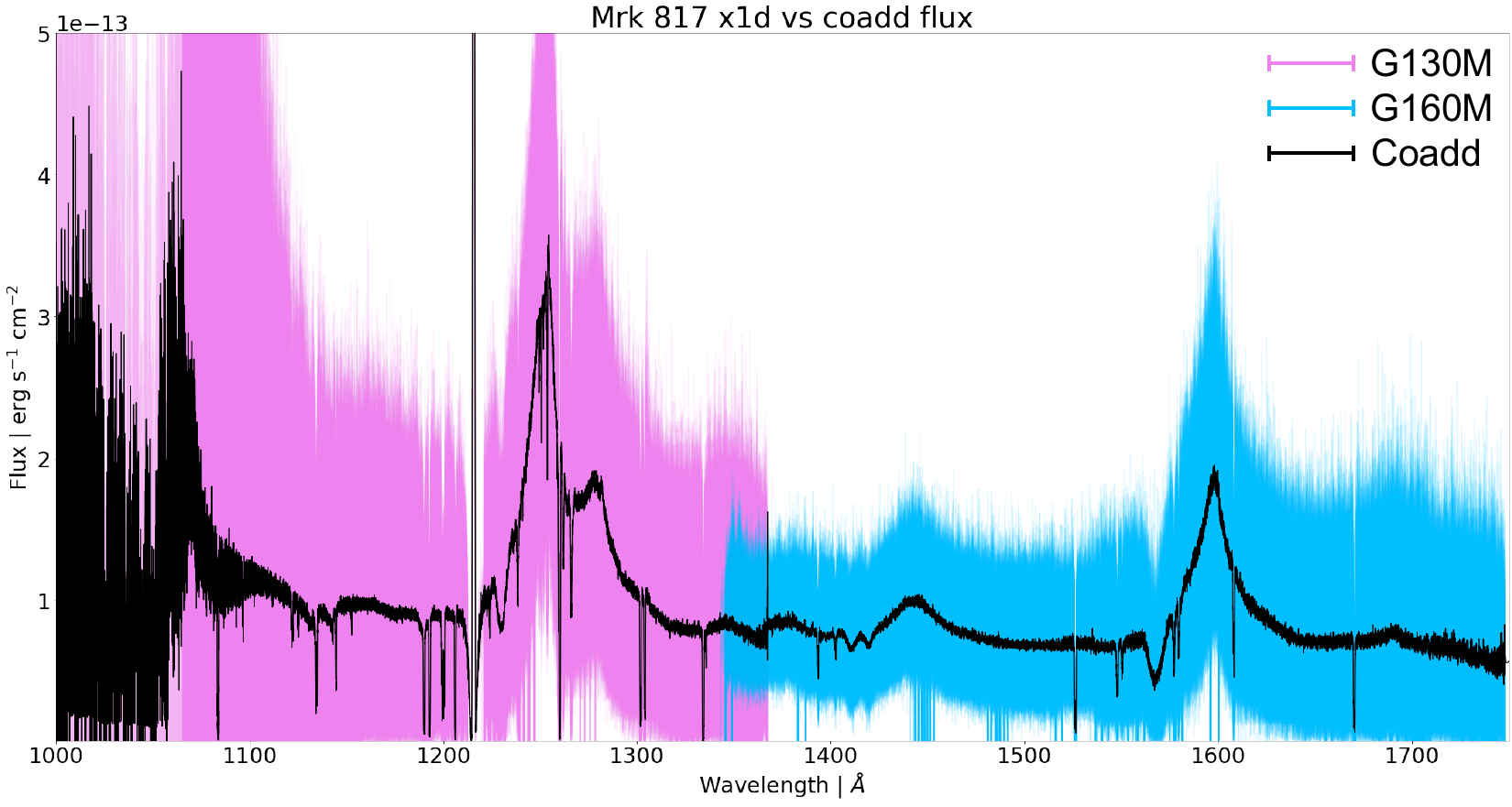}
\vspace{-0.3cm}
\caption{A demonstration of increasing SNR with HASP coadds. 391 G130M and 259 G160M spectra from HST Program ID 16196, which monitored the active galaxy Mrk 817, are plotted in purple and blue, respectively. These spectra form the input for a program-level coadd. The black curve shows the coadded and abutted COS G130M/G160M program-level product. Individual spectra produce an average signal-to-noise ratio $\sim$3 at 1200~\AA, and the program-level coadd produces a signal-to-noise ratio of 56.8. Note that the original program consisted of 676 input G130M  and 634 G160M spectra, but due to the variable nature of the target many spectra were filtered by the flux checking algorithm.}
\label{fig:mrk817}
\end{figure*}

We present two illustrative examples of cases where HASP coadds provide additional scientific benefit over default archive products. In general, the automated default HASP coadds give users immediate access to a) automatic combining of different CENWAVES within a grating, b) higher SNR program coadds compared to their individual exposures, and c) significant wavelength coverage through the abutment of different gratings. An example of increased SNR is apparent in the case of GO 16196 (PI: Kriss; See Figure \ref{fig:mrk817}), which obtained 391 and 259 spectra for COS/G130M and COS/G160M respectively as part of a long term monitoring of an active galaxy. Individual input x1d spectra had SNR of $\sim3$ at 1200~\AA, but the HASP coadd produces an SNR of 56 at the same wavelength. 

The SNR estimate is derived from a combination of the estimated uncertainties per wavelength bin for all input files and should be checked by the user if it exceeds the typical limiting SNR for a given grating. The SNR may be overestimated in the limit of many combinations of the same grating/CENWAVE, but are dependent on the observing details of a given program. Additionally, in the above case, the target is variable, and some spectra may be filtered out by the flux checking algorithm.

\begin{figure*}[!b]
\centering
\includegraphics[width=0.9\textwidth]{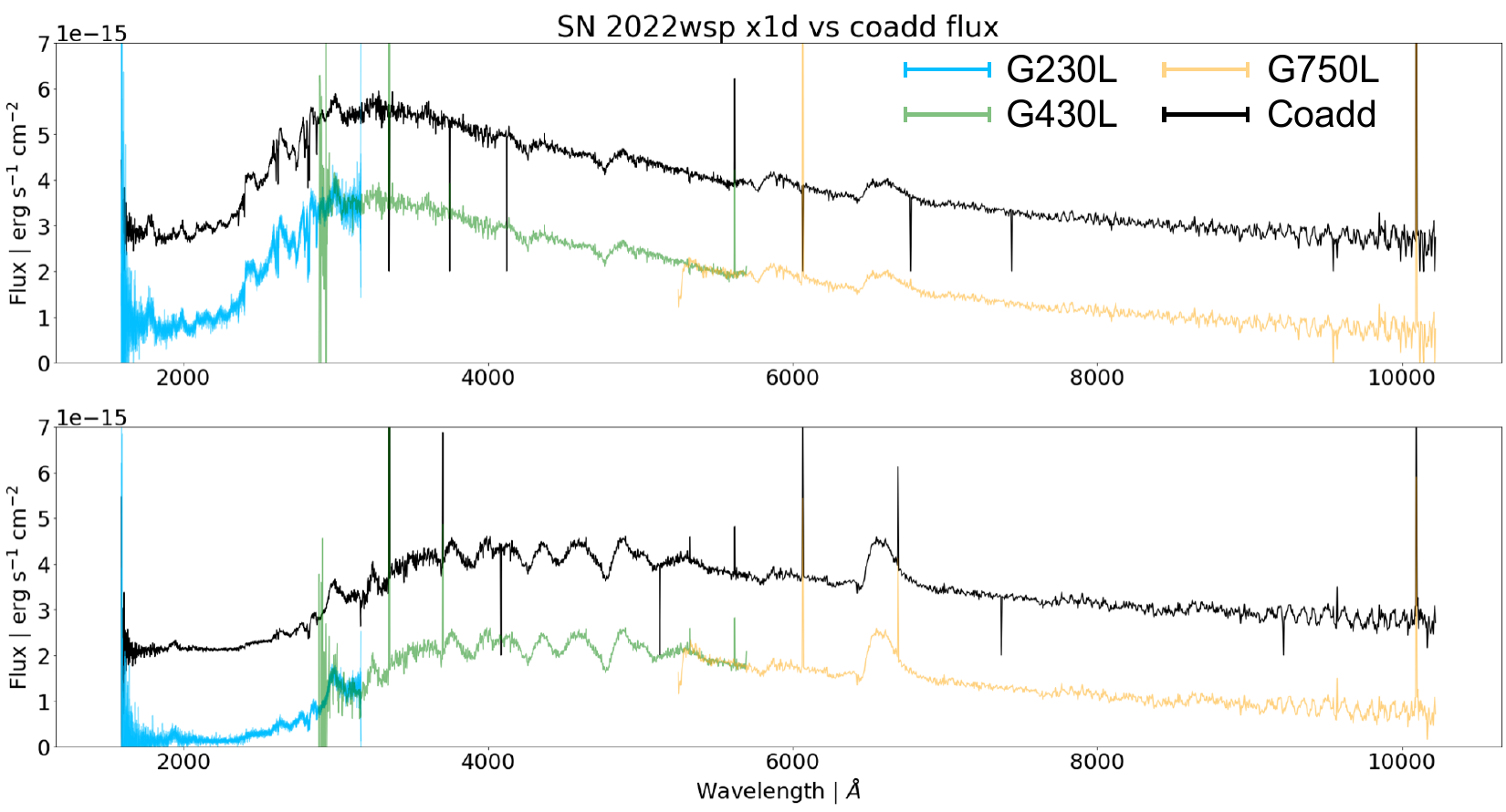}
\vspace{-0.3cm}
\caption{Coadded and abutted STIS G230L/G430L/G750L products from HST Program ID 16656 for supernova 2022wsp in black at +10 days (top panel) and +20 days (bottom panel). G230L, G430L, and G750L constituent spectra that form the coadded products are plotted in blue, green, and orange, respectively. Product fluxes are offset from observed values.
}
\label{fig:snr}
\end{figure*}

Observations of SN2022wsp exemplify the utility of HASP abutments in Figure \ref{fig:snr} (GO 16656; PI: Filippenko). The individual spectra from STIS/G230L, G430L, and G750L produce a low resolution spectral energy distribution of the supernova from 1800~\AA-10000~\AA. The HASP coadd shows a seamless flux distribution across the three gratings. Note also that in this case cosmic rays and CCD fringing (at wavelengths $>$ 7000~\AA) are not removed, and would require further custom processing by the user and a recreation of a coadd using the publicly available \texttt{coadd} code.

Along with these data products, users will be able to download a log file that contains diagnostic information about the coadds, including which files were fed into \texttt{coadd} from MAST, which files were rejected by \texttt{coadd}'s prefiltering and flux checking, and other output and warnings from the coadding procedure as the script creates the visit- and program-level data products. These log files are produced for each PID and will have filenames in the format \texttt{HASP\_<PID>.out}. They are text files that can be opened and viewed in any text editor. An example of an output log with specific features pointed out is shown in Appendix C. 

\vspace{-0.4cm}
\ssection{HASP Testing and Scientific Validation}\label{sec:test}

\begin{figure*}[!ht]
\centering
\includegraphics[width=0.98\textwidth]{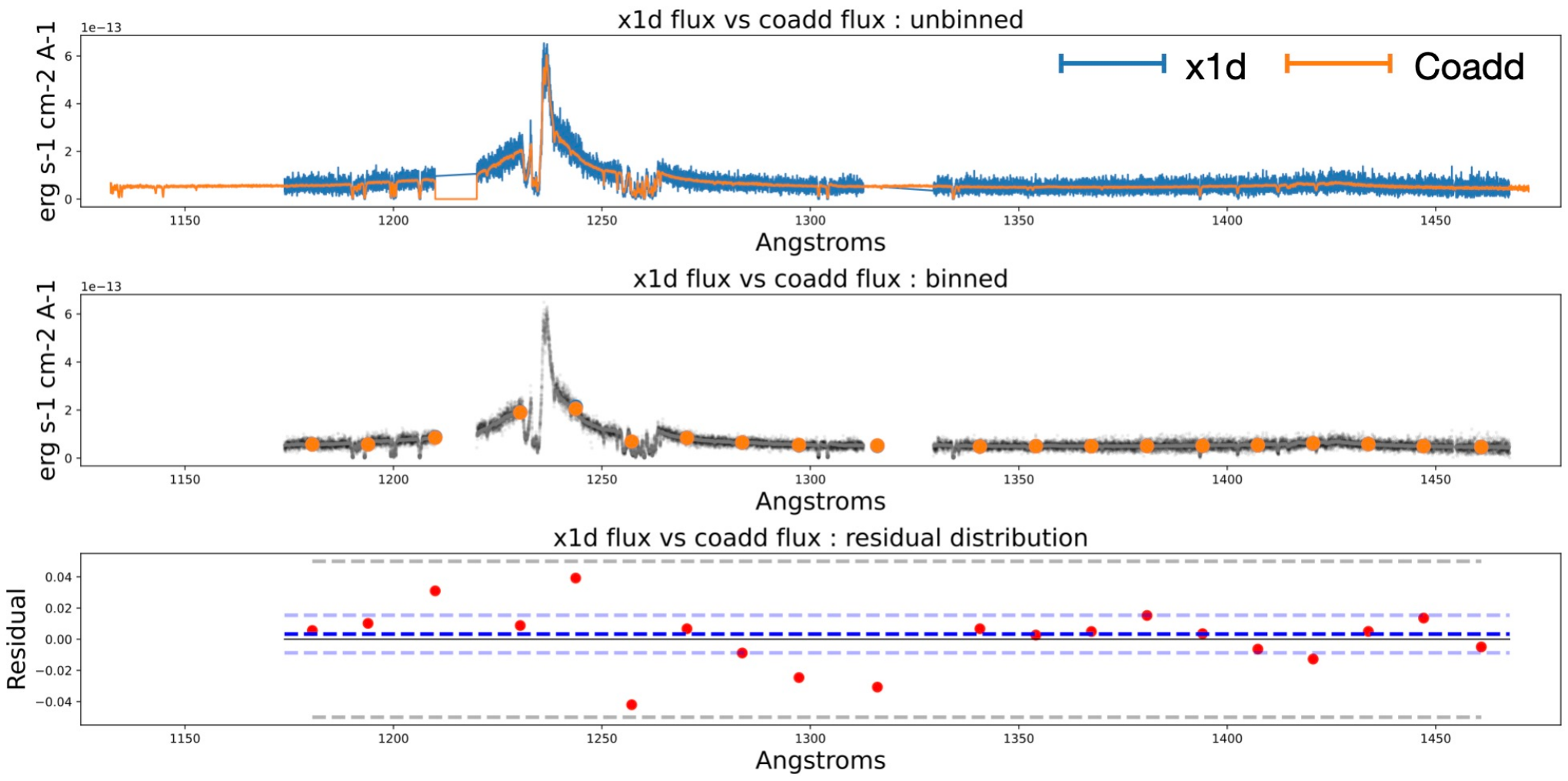}
\vspace{-0.3cm}
\caption{Example flux testing for HASP coadd products. Row 1: Direct comparison between G130M HASP coadd product (orange) and a constituent G130M x1d spectrum (blue) for NGC 5548 (PID: 13330). x1d fluxes are interpolated onto the coadd product wavelength array for direct comparison. Row 2: Individual datapoints for the coadd product and x1d spectrum are in gray. Flux arrays for both datasets are combined into 20 bins, shown as orange and blue circles for the coadd product and x1d spectrum, respectively. As their fluxes are relatively similar, blue circles are often plotted underneath orange circles. Row 3: Residuals between the two binned datasets, using the metric of (x1d - coadd) / x1d. The dark blue dashed line represents the residual mean and the light blue dashed lines represent the residual standard deviation.  The grey dashed lines are the HASP requirements of 5\%. 
}
\label{fig:testing}
\end{figure*}

To ensure the efficacy of our new data products in enhancing individual spectra without introducing degradation, we conduct a comprehensive series of tests comparing each coadd product with its underlying x1d datasets, as illustrated in Figure \ref{fig:testing}. Our initial focus is on examining the agreement between coadd and constituent fluxes. To facilitate a direct flux comparison, x1d fluxes are interpolated to the wavelength array of the coadd dataset. The arrays are filtered for wavelength bins with serious data quality flags. These truncated arrays are then divided into 20 bins, and we calculate the residuals between datasets using the formula \( r = (\text{{F$_{\mathrm{x1d}}$}} - \text{{F$_{\mathrm{coadd}}$}}) / \text{{F$_{\mathrm{coadd}}$}} \). Our criterion for success is to have both the median residual and residual standard deviation within 5\%.

\begin{figure*}[!ht]
\centering
\includegraphics[width=0.98\textwidth]{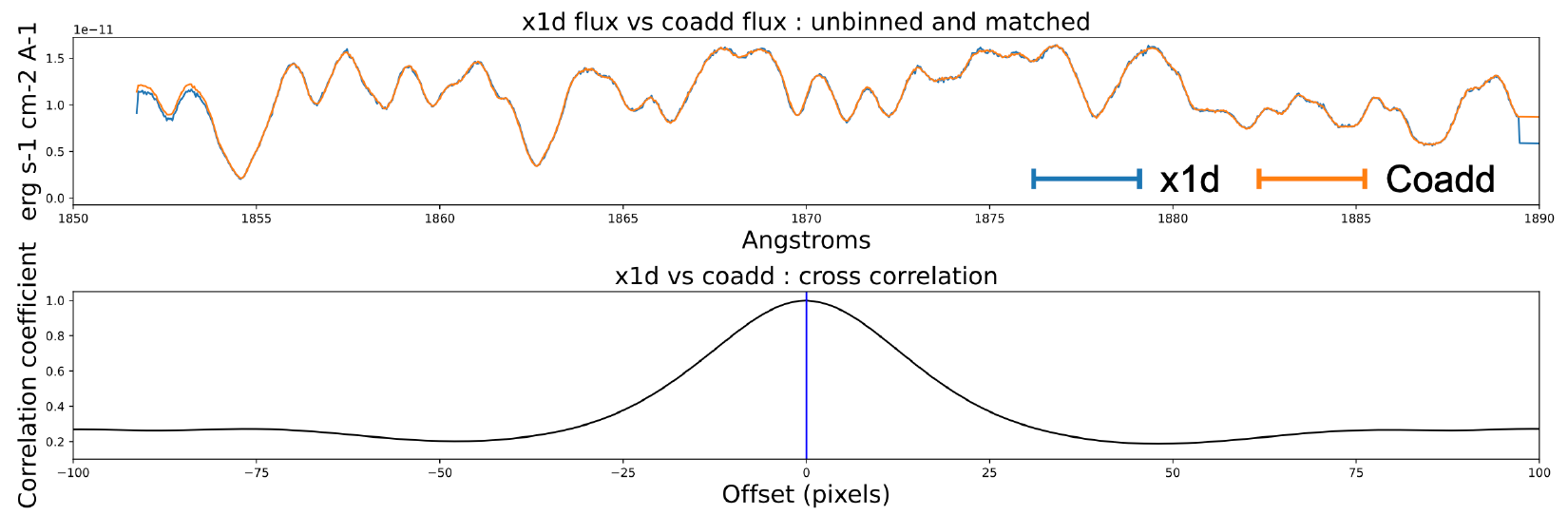}
\vspace{-0.3cm}
\caption{Example wavelength testing for HASP coadd products. Row 1: Direct comparison between G185M HASP coadd product (orange) and a constituent G185M x1d spectrum (blue) for a B star in NGC 3293 (PID: 12520). x1d fluxes are interpolated onto the coadd product wavelength array for direct comparison. Row 2: Correlation coefficient distribution from cross-correlating the coadd product against the x1d constituent spectrum. The blue vertical line represents the peak of the distribution, illustrating that the wavelength offset between coadd and x1d datasets is zero pixels. 
}
\label{fig:testing_wave}
\end{figure*} 

We further examine the alignment of wavelength arrays between datasets to confirm the absence of offsets in the coadd products. Using the previously truncated coadd and x1d datasets, we conduct a cross-correlation analysis, identifying the position of the peak correlation coefficient value in pixels, as depicted in Figure \ref{fig:testing_wave}. Converting this pixel offset to a physical scale, we consider the specific dispersion relation for each instrument/grating setup. The criterion for success is set to be within $\pm$ 1 pixel of a precise match.

Tests to ensure compliance with accuracy requirements were performed across the entire COS and STIS archives, and the results are reported in Table \ref{tab:testing_results}. Flux tests specifically focused on comparisons between coadds and constituent x1d spectra, with a requirement that at least 5 flux bins should have SNR greater than 20. This reduced the size of the original testing archival sample by approximately 22\%. The majority of the datasets that were not used for testing typically had very low SNR across the entire observation in the default extracted 1-D spectra. One example was PID 9066, which obtained 426 STIS 52x2 long slit spectra in parallel to other prime instrument observations. Many spectra intersected blank sky and thus the default 1-D extractions are dominated by cosmic rays rather than any signal.

Using the enhanced SNR sample, we found that the residual mean and standard deviation were both within 5\% for approximately 91\% of x1d datasets. Wavelength tests were carried out on the complete archive samples and demonstrated an offset less than $\pm$ 1 pixel in 97\% of comparisons. Notably, both flux and wavelength tests substantially exceed the benchmark requirement success rate of 75\%, without excluding known extended or variable objects.

\begin{figure*}[!ht]
\centering
\includegraphics[width=0.98\textwidth]{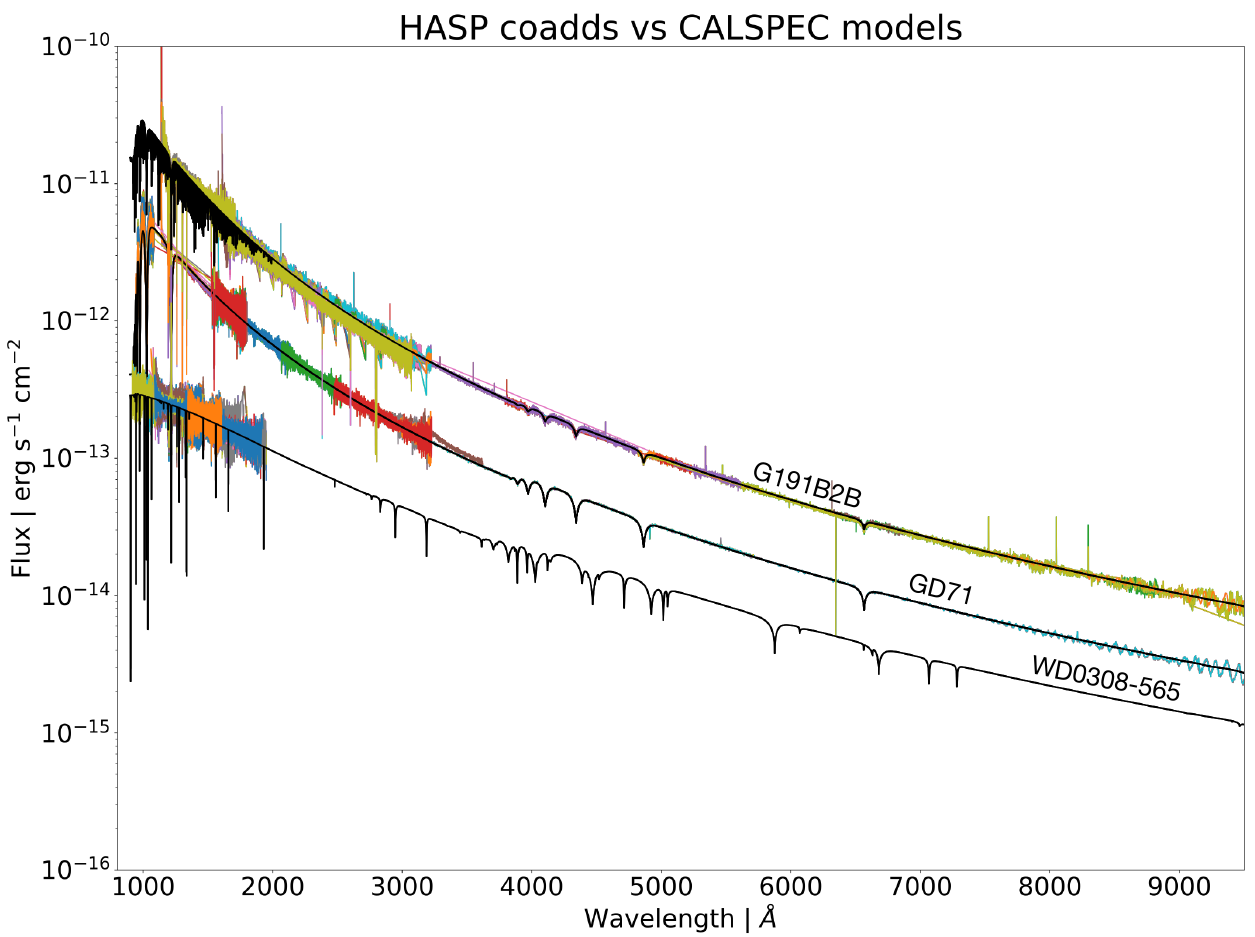}
\vspace{-0.3cm}
\caption{Comparisons between HASP coadd products (colored distributions) and CALSPEC models (black) for three HST flux standard stars; GD71, G191B2B, and WD0308-565.
}
\label{fig:calspec}
\end{figure*}

Additionally, our data demonstrates exceptional accuracy in both absolute flux and wavelength calibration. Rigorous testing involved comparing x1d and coadd products and evaluating coadd products against CALSPEC models for several spectrophotometric standard white dwarfs. Figure \ref{fig:calspec} illustrates these comparisons for WD0308, GD71, and G191B2B, showcasing the alignment between coadd products and CALSPEC models. Assessment of flux and wavelength accuracy against these models reveals a remarkable consistency, achieving a match with CALSPEC's absolute flux across the entire wavelength spectrum to better than 0.032\%, with a standard deviation of approximately 3\%.

\begin{deluxetable}{lcccc}
    \tabletypesize{\small}
    \tablewidth{0pt}
    \tablecaption{
    HASP Testing Results
        \label{tab:testing_results}
}
    \tablehead{
    \colhead{Grating} & \colhead{x1d vs coadd} & \colhead{Sufficient} & \colhead{Flux} & \colhead{Wavelength} \\
    & \colhead{Comparisons} & \colhead{Bins\tablenotemark{a}} & \colhead{Success} & \colhead{Success}
    \\
   
    }
 \startdata
{\bf\underline{COS}} & & & & \\
G130M & 22679 & 75.90\% & 93.94\% & 98.94\% \\
G160M & 16082 & 76.54\% & 93.82\% & 99.84\% \\
G140L & 7882 & 55.89\% & 91.56\% & 99.61\% \\
G185M & 2609 & 63.74\% & 97.71\% & 99.54\% \\
G225M & 1074 & 64.06\% & 96.95\% & 99.87\% \\
G285M & 618 & 92.72\% & 94.76\% & 98.38\% \\
G230L & 2233 & 70.53\% & 89.40\% & 99.51\% \\
\hline
{\bf\underline{STIS}} & & & & \\
G140M & 1784 & 50.34\% & 91.65\% & 89.02\% \\
G140L & 4350 & 73.79\% & 89.88\% & 95.54\% \\
G230L & 4031 & 80.75\% & 92.47\% & 91.91\% \\
G230LB & 1287 & 99.07\% & 93.96\% & 98.24\% \\
G230M & 599 & 78.80\% & 91.10\% & 99.55\% \\
G230MB & 721 & 97.92\% & 96.74\% & 97.82\% \\
G430L & 3771 & 96.74\% & 87.17\% & 94.06\% \\
G430M & 2335 & 97.39\% & 88.61\% & 95.33\% \\
G750L & 3092 & 81.08\% & 88.87\% & 90.79\% \\
G750M & 2260 & 94.03\% & 79.62\% & 92.77\% \\
E140M & 4400 & 89.93\% & 88.17\% & 86.78\% \\
E140H & 1778 & 95.89\% & 92.90\% & 99.84\% \\
E230H & 2510 & 95.66\% & 89.46\% & 99.37\% \\
E230M & 3066 & 88.06\% & 93.22\% & 94.05\% \\
\hline
{\bf TOTAL}\tablenotemark{b} & 89161 & 78.01\% & 91.83\% & 97.04\% \\

\enddata
\tablenotetext{a}{Flux tests were performed on x1ds with at least 5 bins with SNRs$>$20.}
\tablenotetext{b}{Reported percentage values are averages.}
\end{deluxetable}

\vspace{-0.4cm}
\ssection{The \texttt{Coadd} Scripts and Custom Coadditions}\label{sec:custom}

The \texttt{coadd} code and its HASP wrapper are available for public use in Space Telescope's GitHub Organization; \texttt{coadd} is located in the ULLYSES repository\footnote{\href{https://github.com/spacetelescope/ullyses/}{https://github.com/spacetelescope/ullyses/}}, and the HASP wrapper for \texttt{coadd} is in the HASP repository\footnote{\href{https://github.com/spacetelescope/hasp}{https://github.com/spacetelescope/hasp}}. 
Both codes are scripted in Python (compatible with versions 3.9-3.11) and are installable using \href{https://pypi.org/project/pip/}{pip}. More detailed installation and use instructions are given in our tutorial notebooks, which are described below. 

In recognition that users may wish to customize how they coadd spectra, the HASP team has created several user-friendly Jupyter notebooks that demonstrate how to use the HASP scripts beyond the default process. These notebooks guide users through several custom processing examples, such as how to correct for slight flux variations between datasets by flux scaling, how to change the flux filtering threshold or turn off the filtering completely, and how to coadd datasets that were rejected by the pre-filter.

The HASP customization goal is to allow users to fine tune the coaddition process according to their specific requirements. By giving users a comprehensive understanding of the underlying processes via easy-to-use tutorials, we hope to facilitate informed decision-making in the custom coadd creation process. This flexibility and transparency will enhance the project's accessibility and utility, catering to diverse research needs and promoting a deeper exploration of HST's spectral data.

The following notebooks are being released along with the HASP data products:

\noindent\textbf{Setup} guides users through the basics of setting up and running \texttt{coadd}. It details installing conda, creating a conda environment, downloading the HASP scripts through GitHub, and finally installing and using the scripts. This tutorial also informs users of the STIS and COS filetypes needed to create coadded spectral products, as well as explains the output data product file naming convention. Lastly, it shows how to run \texttt{coadd} with different qualifier flags to customize how \texttt{coadd} behaves.

\noindent\textbf{Introduction to \texttt{coadd}} provides users with more detailed examples of running the HASP script on COS and STIS datasets for the standard star, GD71. This notebook shows how to use the Python package \texttt{Astroquery} to download COS and STIS data and run \texttt{coadd} on these files. Users are then shown how to use the Python plotting package \texttt{Matplotlib} to create a plot of the flux and signal-to-noise for the newly coadded spectra. The tutorial also demonstrates the use of the (optional) flux threshold flag for \texttt{coadd}, which can be changed to alter the number/quality of files used in the coaddition.

\noindent\textbf{Data Diagnostics} guides users on how to open and parse the \texttt{coadd} script output logs, which contain useful information about which datasets were used or rejected from the coadd data products and why. It shows users how to turn off the default prefilters, so that any datasets can be used in the final coadd. This tutorial also shows examples of how to assess the quality of rejected datasets, as well as how to turn off the flux filtering completely.

\noindent\textbf{Flux Scaling} goes in depth about checking for variability in a dataset, scaling the data to a common median flux, running \texttt{coadd}, and inspecting the output by plotting flux as a function of wavelength. Because \texttt{coadd}’s flux filter is designed to prevent data from failed observations from being coadded, it can have inadvertent effects on other good quality datasets. Some examples include: variable sources, extended sources observed with different sized apertures or at multiple orientations, and data observed using the small STIS apertures that can be impacted by changes in observatory focus. If a user's science case is not dependent on the accuracy of a dataset's absolute flux, scaling the input spectra to be the same average or median flux may be desirable.

The custom coaddition notebooks will be hosted on the \href{https://spacetelescope.github.io/hst_notebooks/}{Space Telescope HST Notebook Repository HQ}. As time and resources allow, additional notebooks will be delivered with future releases of HASP to help users maximize the science from their preferred spectral datasets.

\vspace{-0.4cm}
\ssection{HASP Caveats}\label{sec:caveats}

HASP products are not designed to meet every science case for every type of astrophysical object. We list the specific caveats below, but users are encouraged to also monitor COS STScI Analysis Newsletters and future Instrument Science Reports (ISR) for any additions to this list. We also detail in this section specific types of targets and science cases that may not meet the typical accuracy requirements for COS and STIS modes. It is always advisable not to blindly trust the coadd results, and to make use of this ISR, the HASP Data Diagnostics Notebook, and other COS and STIS documentation to assess the quality of a default HASP coadd.

The code coadds MAST products produced using the standard CalCOS and CalSTIS pipelines, so if there are custom reductions that would need to be done for standard pipeline products, it will need to be done for the coadd products. Example use cases are extended target flux extraction, defringing of STIS spectra, and cosmic ray removal for data not taken with CR-SPLITS. Targets observed with different apertures, in particular extended targets, or extended targets observed at different spacecraft orientations, may result in different amounts of flux in each aperture and for each instrument, and as a result spectra may be removed from coadd products by the flux checking algorithm if the flux is sufficiently different between apertures. If such flux differences occur between gratings, there is no check for consistency between modes. Similarly, for variable targets, e.g., AGN or variable stars, the flux checking algorithm may remove exposures for these targets, and coadd products may not be produced. In other cases, the brightest spectra will be retained for a HASP coadd, and if the other spectra are fainter than the flux checking criteria they will be removed. Extended and variable targets were included by definition in our full archive testing and with our existing methodology do not significantly impact our success criteria. As mentioned in Section~\ref{filtering}, moving targets (i.e., Solar system objects) are not coadded at the program level due to their expected variability from visit to visit.

As mentioned in Section \ref{sec:dp}, programs that obtained very high SNR through multiple observations may return estimated SNRs that are higher than the limiting SNR for a given mode. For example, the COS FUV modes have limiting SNRs due to fixed-pattern noise (Roman-Duval et al., 2023), and STIS CCD spectra beyond 7000~\AA\ suffer from fringing effects that need to be removed separately. Users are strongly encouraged to verify that the SNR estimate in a HASP coadd is correct through empirical methods, such as by dividing the mean continuum flux by the standard deviation in the continuum of a source.

\vspace{-0.4cm}
\ssection{Conclusions}\label{sec:conclusions}

The HASP project provides science quality coadded spectra in MAST for each target at the visit and program level in an automated fashion for the first time for COS and STIS.  The archival data used by HASP are meticulously filtered to remove suspect observations resulting in reliable, high-quality combined products. Additionally, HASP provides abutted spectra for quick-look purposes that show the overall wavelength coverage and general data quality for each target in a program.  A key strength of the project is that the publicly available \texttt{coadd} code is readily modified to fine-tune the selection of input spectra as well as the parameters used in the coaddition.  Several custom coaddition examples in Jupyter notebooks have been created by the HASP team and made available as part of the project.

The HASP project is the first step of a larger initiative to create a new, automated and updated HSLA.  The codes and processes developed as part of HASP will directly inform our approach in the next step of this process: to create cross-program target level coadds in a semi-automated way.  HASP will continue as its own project providing combined visit and program level spectra for each target as new data are obtained or additional calibrations for instruments are enacted in the current and upcoming cycles.  There will be continuity and synergy between HASP and the new HSLA as we enhance the legacy value of COS and STIS spectroscopy across the operational lifetime of both instruments for years to come.






\vspace{-0.3cm}
\ssectionstar{Change History for COS ISR 2024-01}\label{sec:History}
\vspace{-0.3cm}
Version 1: 03 January 2024 - Original Document \\
Version 2: 16 January 2024 - Updates to Figure \ref{fig:coaddcenwave} and Appendix C

\vspace{-0.3cm}
\ssectionstar{References}\label{sec:References}
\vspace{-0.3cm}

\noindent
Bauer, Bennet, \& Brown, 2007, ApJS, 171, 249 \\
Peeples et al., 2017, COS Instrument Science Report 2017-4 \\
Roman-Duval et al., 2020, Res. Notes AAS, 4, 205 \\
Roman-Duval et al., 2023, COS Instrument Science Report 2023-11 \\
\newpage
\vspace{-0.3cm}
\ssectionstar{Appendix A: Details of Abutment Algorithm}\label{sec:appendix}
\vspace{-0.3cm}

As described in Section \ref{abutment}, the single grating coadded spectra were combined using an abutment
algorithm.  No attempt was made to optimize signal-to-noise or resolution at each wavelength, instead
a simple "all-or-nothing" approach was taken whereby all the data from one grating are used until a certain
wavelength, and thereafter data from a different grating are used until the next grating transition wavelength.
How are transition wavelengths defined and data from individual gratings prioritized?

The grating priority list and range of use of each grating are governed by the grating priority table, listed in
Table \ref{tab:abutment_priority}.  The wavelength range defines the maximum range over which the data for that grating are
considered valid.  In practice, the actual data from that grating will have a wavelength range that depends on the
CENWAVE (and FP-POS for COS data).  The wavelength range assigned to a single grating spectrum is the
wavelength range of the data that fits in the range defined in the grating priority table.  So, for example,
if the data for a certain COS G160M data product goes from 1340 \AA{} to 1750 \AA, and the grating priority
table lists the wavelength range as 1342--1800 \AA, then the wavelength range used will be 1342--1750 \AA.
These two wavelengths define the transition wavelengths for that grating product, and the relative priorities
for the gratings are determined from the ordering in the grating priority table.

The abutment algorithm is then simply deciding, at each of these transition wavelengths, which grating
data has the highest priority, and, if it is different from the grating currently in use, then switching
to using the higher priority grating.  There are many possible ways of implementing this process; in
practice a method was chosen akin to setting priority bits to each grating, with the bit values ordered by
grating priority.  As the ordered list of transition wavelengths is traversed, if the transition wavelength
corresponds to the lower wavelength limit of a grating, then the bit corresponding to that grating is set,
whereas if we are considering a transition wavelength that is the upper wavelength limit for a grating,
the bit for that grating is unset.
The grating to use starting at each transition wavelength is then the grating corresponding to the highest bit
that is set.

We present an example scenario to illustrate the abutment process.  To start, Table \ref{tab:abutment_data}
lists the data in possession,
including the instrument, grating, and wavelength range for each mode, as well as a bit value arising from
the ordering of the gratings in the grating priority Table \ref{tab:abutment_priority}.  The start and end wavelengths give
us our 8 transition wavelengths, where appropriate bits are set or unset,  depending on whether
the wavelength is the start or end wavelength of a grating's data.  Table \ref{tab:abutment_steps} shows the
effect at each of these transition wavelengths, where we see that the resultant abutted product will be
constructed from the COS/G130M spectrum in the wavelength range of 900--1144 \AA, will then switch to
the STIS/E140M spectrum in range of 1144--1710 \AA, then finally switch to the COS/G160M spectrum in the
range of 1710--1774 \AA. In this case, data from the STIS/E140H spectrum is not used in the final spectral
product, as there is no wavelength range over which it is determined to be the highest priority for this
set of gratings.

\clearpage

\begin{deluxetable}{cccc}
    \tabletypesize{\small}
    \tablewidth{0pt}
    \tablecaption{
    Example dataset for abutment demonstration.\label{tab:abutment_data}}
    \tablehead{
    \colhead{Instrument}&\colhead{Grating}&\colhead{Wavelength Range}&\colhead{Priority Bit Value}
    \\
    & & \colhead{[\AA]} & 
    }
    \startdata
    STIS & E140M & 1144 -- 1710 & 1000 \\
    COS & G130M & 900 -- 1450 & 0100 \\
    COS & G160M & 1360 -- 1775 & 0010 \\
    STIS & E140H & 1139 -- 1700 & 0001 \\
    \enddata
\end{deluxetable}
\vspace{-10pt}
\begin{deluxetable}{ccccc}
    \tabletypesize{\small}
    \tablewidth{0pt}
    \tablecaption{
    Example dataset abutment priorities demonstration.     \label{tab:abutment_steps}}
    \tablehead{
    \colhead{Wavelength}&\colhead{Bit change}&\colhead{Bits Set}&\colhead{Highest Bit}&\colhead{Result}
    \\
    \colhead{[\AA]} & & & \colhead{Set} &
    }
    \startdata
    900 & Set 0100 & 0100 & 0100 & begin with COS/G130M \\
    1139 & Set 0001 & 0101 & 0100 & keep using COS/G130M \\
    1144 & Set 1000 & 1101 & 1000 & switch to STIS/E140M \\
    1360 & Set 0010 & 1111 & 1000 & keep using STIS/E140M \\
    1450 & Unset 0100 & 1011 & 1000 & keep using STIS/E140M \\
    1700 & Unset 0001 & 1010 & 1000 & keep using STIS/E140M \\
    1710 & Unset 1000 & 0010 & 0010 & switch to COS/G160M \\
    1775 & Unset 0010 & 0000 & 0000 & end with COS/G160M \\
    \enddata
\end{deluxetable}

\vspace{-2.6cm}
\ssectionstar{Appendix B: Structure of HASP BINTABLE Extensions}\label{sec:appendixb}
\vspace{-0.3cm}

The HASP BINTABLE consists of a science extension containing the wavelength, flux, error, estimated SNR, and effective exposure time for each wavelength bin in the coadd. Table \ref{tab:datacolumns} provides the name for each column in the science extension, its units and a brief description of the column.

Additionally, the other extension in a HASP coadd BINTABLE details the provenance of the product, and lists the details of each input spectrum. Users can parse this information, along with the HASP logfile (See Appendix C for details) when determining if any input files were filtered, or to understand the original observing details of a given input dataset. Table \ref{tab:provenance} gives the detailed list of the column names, the units if they are not ascii characters, and a brief description.

Users should note that in order to be compatible with the ULLYSES HLSPs, the provenance table includes a column that denotes the observatory that generated the spectrum (TELESCOPE). In the case of HASP, only HST spectra will be included.

\clearpage

\begin{deluxetable}{lll}
    \tabletypesize{\small}
    \tablewidth{0pt}
    \tablecaption{
    Columns available in a HASP coadd BINTABLE Science Extension. 
    \label{tab:datacolumns}} 

\tablehead{
    \colhead{Keyword} & \colhead{Units} & \colhead{Description}}
\startdata
WAVELENGTH & \AA & Wavelength bin \\
FLUX & ergs cm$^{-2}$ s$^{-1}$ \AA$^{-1}$ & Coadd flux \\
ERROR & ergs cm$^{-2}$ s$^{-1}$ \AA$^{-1}$ & Uncertainty on the flux \\
SNR & ...  & Estimate of the SNR per wavelength bin \\
EFF\_EXPTIME & s & Weighted exposure time of input flux bins \\
\enddata

\end{deluxetable}

\begin{deluxetable}{lll}
    \tabletypesize{\small}
    \tablewidth{0pt}
    \tablecaption{
    Columns available in a HASP coadd BINTABLE Provenance Table. 
    \label{tab:provenance}} 

\tablehead{
    \colhead{Keyword} & \colhead{Units} & \colhead{Description}}
\startdata
FILENAME & ...  & Input Spectrum Filename\\
EXPNAME & ...  & Exposure name, if multiple spectra per file \\
PROPOSID & ... & Proposal ID \\
TELESCOPE & ... & Observatory \\
INSTRUMENT & ...  & Instrument \\
DETECTOR & ...  & Instrument Detector\\
DISPERSER & ... & Grating \\
CENWAVE & ... & Central Wavelength of Grating\\
APERTURE & ... & Aperture Selected \\
SPECRES & ...  & Estimated Spectral Resolution \\
CAL\_VER & ... & Calibration Version \\
MJD\_BEG & d & Exposure Start Time (MJD) \\
MJD\_MID & d & Exposure Mid-Point (MJD) \\
MJD\_END & d & Exposure End Time (MJD) \\
XPOSURE & s  & Exposure Time \\
MINWAVE & \AA & Minimum wavelength\\
MAXWAVE & \AA & Maximum wavelength \\
\enddata

\end{deluxetable}

\clearpage

\vspace{-0.3cm}
\ssectionstar{Appendix C: Example of a HASP Trailer File}\label{sec:appendixc}
\vspace{-0.3cm}

Below is an example trailer file for the GO program 13784. Text in black is directly from the log, and text in \textcolor{cyan}{cyan} is added for explanation. \\

{\footnotesize
\noindent\textcolor{cyan}{The top lines of the trailer that begin with a timestamp and ``INFO'' are output from the poller. It records if any of the eligible files from the initial program query (e.g. \texttt{x1d} and \texttt{sx1} files) were removed due to data quality issues (like those with AlertObs) or those in the SkipAlignment table (which includes special calibration programs, statically archived datasets, etc.). In this example, we see that the COS file  lclb05iyq was removed from the coaddition.}

\begin{verbatim}
2024004140317 INFO src=hasp_retrieve_inputs.filter_out_low_quality_inputs
fsn=clb msg="All input_files met the quality requirements"

2024004140317 INFO src=hasp_retrieve_inputs-hap_verify_poller
_file.remove_entries_to_skip msg="Checking whether any ipppssoots
are in the SkipAlignment table"

2024004140317 INFO src=hasp_retrieve_inputs-hap_verify_poller
_file.remove_entries_to_skip msg="Removed 1 entries from the poller
file because these were found in the SkipAlignment table: ['lclb05iyq']"

2024004140323 INFO src=hasp_coadds.get_resource_path_values msg="Using 
haspsp.resource and repro_sci.path"

2024004140323 INFO src=hasp_coadds-run_swrapper.run_swrapper_script 
fsn=clb msg="Running swrapper -i ...clb -o ...clb in the calibration 
software conda environment: caldp_20231208"
\end{verbatim}

\noindent\textcolor{cyan}{The following the text is generated by \texttt{coadd} or its wrapper. First, the code checks the remaining input files for any qualities to filter out. This program has no additional rejected datasets, but if it did, a line with the dataset name and the reason for removal will be printed in the log, for example, ``File ...o57d02020\_x1d.fits removed from products because FGSLOCK = FINE/GYRO''. After the prefiltering is complete, the remaining files are sorted and listed. Note that some of these files may later be removed via the flux filter.}

\begin{verbatim}
HASP version 0.9.5
Creating list of unique modes from these files:
...lclb05iyq_x1d.fits LSQ15ABL COS FUV G140L PSA 13784 (13784, '05')
...lclb05j0q_x1d.fits LSQ15ABL COS FUV G140L PSA 13784 (13784, '05')
...lclb05j2q_x1d.fits LSQ15ABL COS FUV G140L PSA 13784 (13784, '05')
...lclb05j4q_x1d.fits LSQ15ABL COS FUV G140L PSA 13784 (13784, '05')
...lclb05j6q_x1d.fits LSQ15ABL COS FUV G140L PSA 13784 (13784, '05')
...lclb05jfq_x1d.fits LSQ15ABL COS FUV G140L PSA 13784 (13784, '05')
...lclb05k6q_x1d.fits LSQ15ABL COS FUV G140L PSA 13784 (13784, '05')
...lclb05kbq_x1d.fits LSQ15ABL COS FUV G140L PSA 13784 (13784, '05')
...oclb06010_x1d.fits LSQ15ABL STIS NUV-MAMA G230L 52X0.2 13784 (13784, '06')
...oclb06020_x1d.fits LSQ15ABL STIS NUV-MAMA G230L 52X0.2 13784 (13784, '06')
...oclb06030_x1d.fits LSQ15ABL STIS NUV-MAMA G230L 52X0.2 13784 (13784, '06')
...oclb06040_x1d.fits LSQ15ABL STIS NUV-MAMA G230L 52X0.2 13784 (13784, '06')
\end{verbatim}

\noindent\textcolor{cyan}{\texttt{coadd} now begins the process of computing visit-level data products. First, each grating in the visit is processed, and a data product is written. Note that if a file is flagged for removal by the flux filter, a warning that says ``Removing file (filename) from product'' will be written after the line ``Using a maximum SNR of 20 in flux-based filtering''. No files are removed in this example.}

\begin{verbatim}
Looping over visits
Processing product (13784, '05')
Targets in visit (13784, '05'): ['LSQ15ABL']
Processing target LSQ15ABL in visit (13784, '05')
Processing grating COS/G140L
Importing files ['...lclb05iyq_x1d.fits', '...lclb05j0q_x1d.fits', 
'...lclb05j2q_x1d.fits', '...lclb05j4q_x1d.fits', '...lclb05j6q_x1d.fits', 
'...lclb05jfq_x1d.fits', '...lclb05k6q_x1d.fits', '...lclb05kbq_x1d.fits']
Processing file ...lclb05iyq_x1d.fits
Processing file ...lclb05j0q_x1d.fits
Processing file ...lclb05j2q_x1d.fits
Processing file ...lclb05j4q_x1d.fits
Processing file ...lclb05j6q_x1d.fits
Processing file ...lclb05jfq_x1d.fits
Processing file ...lclb05k6q_x1d.fits
Processing file ...lclb05kbq_x1d.fits
Using a maximum SNR of 20.0 in flux-based filtering
   Wrote ...hst_13784_cos_lsq15abl_cg140l_lclb05_cspec.fits
No need to create abutted product as < 2 single grating products
\end{verbatim}

\noindent\textcolor{cyan}{If there were more gratings observed in this visit, those would be abutted next. Since there are not in this example, the process starts again for the next visit in the program.}

\begin{verbatim}
Processing product (13784, '06')
Targets in visit (13784, '06'): ['LSQ15ABL']
Processing target LSQ15ABL in visit (13784, '06')
Processing grating STIS/G230L
Importing files ['...oclb06010_x1d.fits', '...oclb06020_x1d.fits', 
'...oclb06030_x1d.fits', '...oclb06040_x1d.fits']
Processing file ...oclb06010_x1d.fits
Processing file ...oclb06020_x1d.fits
Processing file ...oclb06030_x1d.fits
Processing file ...oclb06040_x1d.fits
Using a maximum SNR of 20.0 in flux-based filtering
   Wrote ...hst_13784_stis_lsq15abl_sg230l_oclb06_cspec.fits
No need to create abutted product as < 2 single grating products
\end{verbatim}

\noindent\textcolor{cyan}{\texttt{coadd} now begins the process of computing program-level data products.}
\begin{verbatim}
Looping over proposals
Processing product 13784
Targets in proposal 13784: ['LSQ15ABL']
Processing target LSQ15ABL in proposal 13784
Processing grating COS/G140L
Importing files ['...lclb05iyq_x1d.fits', '...lclb05j0q_x1d.fits', 
'...lclb05j2q_x1d.fits', '...lclb05j4q_x1d.fits', '...lclb05j6q_x1d.fits', 
'...lclb05jfq_x1d.fits', '...lclb05k6q_x1d.fits', '...lclb05kbq_x1d.fits']
Processing file ...lclb05iyq_x1d.fits
Processing file ...lclb05j0q_x1d.fits
Processing file ...lclb05j2q_x1d.fits
Processing file ...lclb05j4q_x1d.fits
Processing file ...lclb05j6q_x1d.fits
Processing file ...lclb05jfq_x1d.fits
Processing file ...lclb05k6q_x1d.fits
Processing file ...lclb05kbq_x1d.fits
Using a maximum SNR of 20.0 in flux-based filtering
   Wrote ...hst_13784_cos_lsq15abl_cg140l_lclb_cspec.fits
Processing grating STIS/G230L
Importing files ['...oclb06010_x1d.fits', '...oclb06020_x1d.fits', 
'...oclb06030_x1d.fits', '...oclb06040_x1d.fits']
Processing file ...oclb06010_x1d.fits
Processing file ...oclb06020_x1d.fits
Processing file ...oclb06030_x1d.fits
Processing file ...oclb06040_x1d.fits
Using a maximum SNR of 20.0 in flux-based filtering
   Wrote ...hst_13784_stis_lsq15abl_sg230l_oclb_cspec.fits
\end{verbatim}

\noindent\textcolor{cyan}{Last, the two gratings in each visit of the program are abutted to make the final data product.}

\begin{verbatim}
Making a product from these gratings
COS/G140L 901-2150 (Actual: 1110.1-2278.5)
STIS/G230L 1582.0-3158.7 (Actual: 1580.0-3143.6)
Transition wavelengths tweaked
Transition wavelengths tweaked
Starting at the short wavelength end with grating COS/G140L
Abutting STIS/G230L product to current result
With a transition wavelength of 2150
Truncating current grating at 3143.619142125499
   Wrote ...hst_13784_cos-stis_lsq15abl_cg140l-sg230l_lclb_cspec.fits
\end{verbatim}}

\end{document}